\documentclass[aip,chaos,reprint]{revtex4-1}

\usepackage{graphicx}
\usepackage{amssymb,amsfonts,amsmath}
\usepackage{epsf}
\usepackage{subfigure}
\usepackage{epstopdf}
\usepackage{dcolumn}
\usepackage{bm}
\usepackage{mathrsfs}
\usepackage{mhchem}
\usepackage{color}
\usepackage{ulem}

\newcommand{\be}{\begin{equation}}
\newcommand{\ee}{\end{equation}}
\newcommand{\bea}{\begin{eqnarray}}
\newcommand{\eea}{\end{eqnarray}}

\begin{document}

\title{\bf Template-directed growth of copolymers}

\author{Pierre Gaspard}
\affiliation{Center for Nonlinear Phenomena and Complex Systems,\\
Universit\'e Libre de Bruxelles (U.L.B.), Code Postal 231, Campus Plaine,
B-1050 Brussels, Belgium}

\begin{abstract}
The theory of multistate template-directed reversible copolymerization is developed by extending
the method based on iterated function systems to matrices, taking into account the possibility of multiple activation states instead of a single one for the growth process.  In this extended theory, the mean growth velocity is obtained with an {\it iterated matrix function system} and the probabilities of copolymer sequences are given by matrix products defined along the template.  The theory allows us to understand the effects of template heterogeneity, which include a fractal distribution of local growth velocities far enough from equilibrium, and a regime of sublinear growth in time close to equilibrium.
\end{abstract}

\vskip 0.5 cm

\maketitle

\section{Introduction}

{\bf Copolymers are macromolecular chains composed of several types of monomeric units and their sequence may encode information, as it is the case for DNA in biological cells.  As a matter of fact, these sequences are generated from a template during nonequilibrium processes of copolymerization.  In the example of DNA replication, a DNA template is copied into a complementary DNA strand by enzymes called polymerases.  As recently discovered, the kinetic equations of such template-directed growth processes can be exactly solved in the long-time limit by iterating rate-dependent mathematical functions along the template sequence.  These methods reveal complex behaviors induced by the heterogeneity of the template and they show that replication errors depend on nonequilibrium conditions.}

Growth processes are ubiquitous nonequilibrium phenomena.\cite{L80,G92,M98,P04}  In these processes, atoms or molecules aggregate into larger and larger structures with specific shapes and properties.  In the case of crystal growth, atoms or molecules undergo self-assembly from fluid to solid forming structures that are spatially expanding with a velocity and a shape controlled by the nonequilibrium conditions of their growth.\cite{M95,S96,SMV12,SLV18}  These nonequilibrium processes may generate a large variety of different shapes, as illustrated by snowflakes.\cite{BH62}  In other circumstances, the internal structure of the growing solid may also be determined by the nonequilibrium conditions, as in eutectic growth.\cite{FM92,S96}

In biology, growth processes play an essential role in the morphogenesis of eukaryotic pluricellular organisms, as well as for bacterial colonies or the pluricellular stages in the life cycle of unicellular species such as {\it Dictyostelium} amoebae.\cite{G96} Here, growth is regulated by intercellular communication, influencing the intracellular biochemical reaction networks that control cell division or cell motility.

In between the physical and biological processes, the growth of copolymers is a basic mechanism for the generation of large and complex molecular structures.  Copolymers are macromolecules such as polymers, but composed of different species of monomers.  They grow by the formation of covalent bonds between the monomers.  These copolymerization processes may happen without template as in synthesis of rubber or olefin copolymers, or directed by a template as in DNA replication, transcription, and translation, respectively yielding DNA from DNA, RNA from DNA, and polypeptides from RNA.  These fundamental processes of biology are catalyzed by enzymes known as polymerases for the growth of DNA or RNA macromolecules, and by ribosomes for polypeptides.\cite{Alberts}

Template-directed copolymerization is a nonequilibrium process, in which a copolymer is growing on a template.  By this mechanism, the copolymer sequence is closely related to the template sequence if the copying errors occurring at the molecular scale are low enough.  In this regard, the sequence of the growing copolymer may itself depend on the nonequilibrium conditions.
A key issue is thus to determine how the quantities characterizing the process, including the mean growth velocity, the statistical properties of the copolymer sequence, as well as the thermodynamic properties such as the entropy production will vary with the distance from equilibrium.  Template-directed copolymerization can be investigated using chemical kinetic theory and the theory of stochastic processes.\cite{S75,H89,McQ67,S76,NP77,AG08,CQ09,SC12,C13}  The challenge is that the template constitutes a heterogeneous medium for the growth of the copolymer copy.  Yet, this challenge can be overcome as shown in the present paper.  Remarkably, the kinetic equations of template-directed copolymerization can be exactly solved in the long-time limit by using mathematical iterations similar to those introduced in dynamical systems theory and known as {\it iterated function systems}.\cite{BD85}  These iterations may generate fractal distributions for the local growth velocity along heterogeneous templates.  The purpose of the present paper is to show the generality of these methods, as they apply to processes involving single or multiple activation states for copolymerization.  This aspect is important because these activation states can be considered as different structural states for the enzyme catalyzing template-directed copolymerization and polymerases are observed to have several such states.\cite{TJ06}  With this aim, the matrix product method developed in Ref.~\onlinecite{G19JCP} for multistate template-free copolymerization is here extended to multistate template-directed copolymerization, generalizing results previously obtained for single-state template-directed copolymerization in Refs.~\onlinecite{G16PRL,G17JSM,G17PRE,LZSOL19}.  In this context, the use of matrix products finds its origin in pioneering work by Coleman and Fox on the multistate mechanism for irreversible polymerization processes.\cite{CF63JCP,CF63JACS,CF63JPS}  As shown in Ref.~\onlinecite{G19JCP}, the matrix product method can be extended to reversible processes without template.  Here, the matrix product method is combined with iterated function systems to consider template-directed processes.

The plan of the paper is the following.  The general kinetic theory for multistate template-directed copolymerization is presented in Sec.~\ref{Kinetics} where the kinetic equations are exactly solved in terms of iterated function systems generalized to matrices.  In Sec.~\ref{1state}, the theory is shown to reduce to the previous results obtained for single-state template-directed copolymerization yielding copy sequences given by Bernoulli chains.  Section~\ref{Multistate} is specifically devoted to multistate template-directed copolymerization.  These processes are characterized by their mean growth velocity.  Furthermore, the local growth velocity is shown to have fractal distributions in regimes far enough from equilibrium.  In a regime close to equilibrium, the growth is shown to be sublinear instead of linear in time.  These complex behaviors find their origin in the heterogeneity of the template.
The conclusion is drawn in Sec.~\ref{Conclusion}.

\section{General kinetic theory of template-directed copolymerization}
\label{Kinetics}

\subsection{Kinetic equations}

We consider the growth of the copolymer sequence $\omega=m_1m_2\cdots m_{l-1}$ along the infinite template sequence $\alpha=n_1n_2\cdots n_{l-1}n_l n_{l+1}\cdots$.  The copolymer and template sequences are composed of $M$ different species of monomeric units: $m,n\in\{1,2,...,M\}$.
The template is here taken as a Bernoulli chain of probabilities $(\nu_1,\nu_2,...,\nu_M)$ with $\sum_{n=1}^M\nu_n=1$.

The growth proceeds by the attachment of some monomer $m_l\in\{1,2,...,M\}$ coming from the surrounding solution at the successive locations $l$ of the template.  The process is supposed to be reversible, i.e., the eventual detachment of the last monomeric unit at the tip of the copolymer is also possible.  Moreover, the growing tip of the copolymer may undergo transitions $i\to j$ between several activation states $i,j\in\{1,2,...,I\}$.  These transitions may correspond to conformational changes between different structural states of the catalyst.  Accordingly, the kinetic scheme is described by the following reactions,
\bea
 \begin{array}{l}
 \end{array}
&&\quad\quad\ \ \,
\begin{array}{l}
\omega  = m_1m_2\cdots m_{l-1},\ i\\
\alpha  = n_1\ n_2\; \cdots \, n_{l-1}\, n_{l} \ n_{l+1}\ \cdots
\end{array}
 \quad
 \begin{array}{l}
 + \ m_{l} \\
 \end{array}
\nonumber\\
&&\underset{w_{-m_l,l}^{i}}{\overset{w_{+m_l,l}^{i}}{\rightleftharpoons}}
\begin{array}{l}
\omega'=m_1m_2\cdots m_{l-1} m_{l},\ i\\
\alpha\; =n_1\ n_2\; \cdots \, n_{l-1}\, n_{l} \ n_{l+1}\ \cdots
\end{array}
\label{kin_1}
\eea
and
\bea
 \begin{array}{l}
 \end{array}
&&\quad\quad\,
\begin{array}{l}
\omega  = m_1m_2\cdots m_{l-1},\ i\\
\alpha  = n_1\ n_2\; \cdots \, n_{l-1}\, n_{l} \ n_{l+1}\ \cdots
\end{array}
 \quad
 \begin{array}{l}
\\
 \end{array}
\nonumber\\
&&\underset{w_{l}^{j\to i}}{\overset{w_{l}^{i\to j}}{\rightleftharpoons}}
\begin{array}{l}
\omega'=m_1m_2\cdots m_{l-1},\ j\\
\alpha\; =n_1\ n_2\; \cdots \, n_{l-1}\, n_{l} \ n_{l+1}\ \cdots
\end{array}
\label{kin_2}
\eea
where $w_{\pm m_l,l}^{i}$ denote the attachment and detachment rates of the monomeric unit $m_l$ at the location $l$ of the template if the activation state is $i\in\{1,2,...,I\}$, and $w_{l}^{i\to j}$ is the rate of  the transition $i\to j$ between the states $i$ and $j$ also at the location $l$ of the template.
The reactions are evolving under low conversion conditions, i.e., the rates are low enough with respect to the pool of monomers in the surrounding solution in order for the concentrations of monomers and thus the rates to remain invariant during the whole growth process.

The kinetics is described using the theory of stochastic processes in terms of the probability distribution
\be
P_t\left(m_1 \cdots m_l,l,i\right)
\equiv P_t\left(m_1 \cdots m_l,\ i \quad\quad\ \atop n_1\, \cdots \, n_l \, n_{l+1}\cdots\right)
\ee
 to find the copolymer sequence $m_1 \cdots m_l$ of length $l$ and the activation state $i$ at time $t$.
The time evolution of this probability distribution is ruled by the coupled master equations
\bea
&&\frac{d}{dt}\, P_t\left(m_1 \cdots m_l ,l,i\right) \nonumber\\
&&= w_{+m_l,l}^{i} \, P_t\left(m_1 \cdots m_{l-1} ,l-1,i\right)
\nonumber\\
&&+\sum_{m_{l+1}} w_{-m_{l+1}, l+1}^{i}
\, P_t\left(m_1 \cdots m_l m_{l+1} ,l+1,i\right)
\nonumber\\
&&+ \sum_{j(\ne i)} w_{l}^{j\to i} \, P_t\left(m_1 \cdots m_l ,l,j\right)
\nonumber\\
&&- \left(\sum_{m_{l+1}} w_{+m_{l+1},l+1}^{i}
+w_{-m_l,l}^{i} + \sum_{j(\ne i)} w_{l}^{i\to j}\right) \nonumber\\
&&\qquad \times P_t\left(m_1 \cdots m_l ,l,i\right) .
\label{kin_eq}
\eea

At the location $l$, the rates may depend on the monomeric unit $n_l$ with which the monomer $m_l$ will pair upon attachment, or this dependence may be extended to some subsequence such as $n_{l-1}n_ln_{l+1}$ around the location $l$.

\subsection{Method for solving the kinetic equations}
\label{Solve}

The growth of the copolymer $\omega$ is similar to the propagation of a front at the mean position $\langle l\rangle_t$.  The mean growth velocity $v=d\langle l\rangle_t/dt$ is determined by the processes taking place in the vicinity of the growing tip.  In this regard, the following set of probabilities are introduced,
\bea
p_t(l,i) &\equiv& \sum_{m_1\cdots m_l} P_t(m_1\cdots m_{l-2}m_{l-1}m_l,l,i) ,\label{p0} \\
p_t(m_l,l,i) &\equiv& \sum_{m_1\cdots m_{l-1}} P_t(m_1\cdots m_{l-2}m_{l-1}m_l,l,i) ,\label{p1} \\
&\vdots&\nonumber
\eea
Accordingly, the mean length can be obtained as $\langle l\rangle_t=\sum_{l,i} l \, p_t(l,i)$.
We note that all the probabilities~(\ref{p0}), (\ref{p1}),... are related to each other by summing over monomeric units as for $p_t(l,i)=\sum_{m_l}p_t(m_l,l,i)$, as well as for the further probabilities. The time evolution of these probabilities is ruled by a hierarchy of equations that can thus be deduced from the master equation~(\ref{kin_eq}).  

Since all these equations are linear, they can be solved by using the method of Fourier decomposition as in Ref.~\onlinecite{G17JSM}. The general solution is thus expressed as a linear superposition of particular solutions of the form
\bea
&& p_t(m_{l-r+1}\cdots m_{l-1}m_l,l,i) \nonumber\\
&& = \exp(s_q t + \imath ql)\,  g_q(m_{l-r+1}\cdots m_{l-1}m_l,l,i)\, ,
\label{sol}
\eea
where $-\pi < q \leq +\pi$ plays the role of wave number, $s_q$ is the dispersion relation to be determined, and $\imath=\sqrt{-1}$.  

In the following, we suppose for simplicity that there exist two activation states ($I=2$), although the results can be extended to the general situation of an arbitrary number $I$ of activation states.  The following matrices are defined with the different rates:
\be
{\boldsymbol{\mathsf W}}_{0,l} \equiv \left(
\begin{array}{cc}
-w_{l}^{1\to 2} & w_{l}^{2\to 1}\\
w_{l}^{1\to 2} & -w_{l}^{2\to 1}
\end{array}
\right) ,
\label{W0-dfn}
\ee
\be
{\boldsymbol{\mathsf W}}_{\pm m,l} \equiv \left(
\begin{array}{cc}
w_{\pm m,l}^{1} & 0\\
0 & w_{\pm m,l}^{2}
\end{array}
\right) ,
\label{W-dfn}
\ee
and
\be
{\boldsymbol{\mathsf A}}_l \equiv \sum_m {\boldsymbol{\mathsf W}}_{+m,l}\, .
\ee
The particular solution is also expressed in matricial form as
\be
{\boldsymbol{\mathsf G}}_q({\bf m},l) \equiv \left(
\begin{array}{cc}
g_q({\bf m},l,1) & g_q({\bf m},l,1)\\
g_q({\bf m},l,2) & g_q({\bf m},l,2)
\end{array} \right)
\label{G_q-dfn}
\ee
with ${\bf m}=m_{l-r+1}\cdots m_{l-1}m_l$.  Substituting these matrices into the master equation~(\ref{kin_eq}) for the probability~(\ref{p0}), we find the following matricial equation for ${\boldsymbol{\mathsf G}}_q(l)$,
\bea
&& s_q\, {\boldsymbol{\mathsf G}}_q(l) = {\rm e}^{-\imath q} {\boldsymbol{\mathsf A}}_l\cdot{\boldsymbol{\mathsf G}}_q(l-1) \nonumber\\
&&+ {\rm e}^{\imath q} \sum_{m_{l+1}} {\boldsymbol{\mathsf W}}_{-m_{l+1},l+1}\cdot{\boldsymbol{\mathsf G}}_q(m_{l+1},l+1) \nonumber\\
&&- {\boldsymbol{\mathsf A}}_{l+1}\cdot{\boldsymbol{\mathsf G}}_q(l) 
- \sum_{m_{l}} {\boldsymbol{\mathsf W}}_{-m_{l},l}\cdot{\boldsymbol{\mathsf G}}_q(m_{l},l) \nonumber\\
&& + {\boldsymbol{\mathsf W}}_{0,l}\cdot{\boldsymbol{\mathsf G}}_q(l) \, .
\label{G_q-eq}
\eea
Further equations can be obtained in the same way for ${\boldsymbol{\mathsf G}}_q({\bf m},l)$ with ${\bf m}=m_l,m_{l-1}m_l,...$.

For growth processes, the dispersion relation $s_q$ is expected to have the form
\be
s_q = - \imath \, v \, q -{\cal D} \, q^2 + O(q^3) \, ,
\label{dispersion1}
\ee
where $v$ is the mean growth velocity of the copolymer chain counted in monomers per second, and $\cal D$ is the diffusivity of the front around its mean drift.  The solution can be similarly expanded in powers of the wave number:
\be
{\boldsymbol{\mathsf G}}_q({\bf m},l) = {\boldsymbol{\mathsf G}}_0({\bf m},l) + q\, \frac{d{\boldsymbol{\mathsf G}}_0}{dq}({\bf m},l) + O(q^2)\, .
\label{G_q-expansion}
\ee

\subsection{The solution in the long-time limit}

In the long-time limit, the probability distribution $p_t(l,i)$ is becoming broader and broader because of the diffusivity, so that the dominant values of the wave number $q$ should take smaller and smaller values.  Consequently, the solution can be obtained in the long-time limit by expanding in powers of the wave number $q$ and considering the limit $q\to 0$.  The zero wave-number limit of the solution is denoted
\be
\pmb{\Psi}_l({\bf m}) \equiv {\boldsymbol{\mathsf G}}_0({\bf m},l) = \left(
\begin{array}{cc}
\psi_l({\bf m},1) & \psi_l({\bf m},1)\\
\psi_l({\bf m},2) & \psi_l({\bf m},2)
\end{array} \right) \label{Psi-dfn}
\ee
with
\be
\psi_l({\bf m},i)= g_0({\bf m},l,i) \, .
\label{psi}
\ee
The purpose of duplicating the columns in the matrices~(\ref{Psi-dfn}) [and previously in Eq.~(\ref{G_q-dfn})] is to provide the normalization of the solution by taking the trace of the matrices according to
\be
{\rm tr}\, \pmb{\Psi}_l({\bf m})= \sum_{i=1}^{2} \psi_l({\bf m},i)\, .
\ee

Since the probabilities~(\ref{p0}), (\ref{p1}),... are interrelated by summing over monomeric units, we should have that
\bea
\sum_{m_l} \pmb{\Psi}_l(m_l)  &=& \pmb{\Psi}_l \, , \label{sum-Psi-1}\\
\sum_{m_{l-1}} \pmb{\Psi}_l(m_{l-1}m_l)  &=& \pmb{\Psi}_l(m_l) \, , \label{sum-Psi-2}\\
&\vdots& \nonumber
\eea
In the long-time limit, the probability to find the growing copolymer at the location $l$ with the monomeric subsequence ${\bf m}=m_{l-r+1}\cdots m_{l-1}m_l$ in the activation state $i$ is given by
\be
\mu({\bf m},l,i) = \frac{\psi_l({\bf m},i)}{{\rm tr}\, \pmb{\Psi}_l} \, , \label{mu-Psi-i}
\ee
and in any activation state by
\be
\mu({\bf m},l) = \sum_{i=1}^2 \mu({\bf m},l,i) = \frac{{\rm tr}\, \pmb{\Psi}_l({\bf m})}{{\rm tr}\, \pmb{\Psi}_l}\, .\label{mu-Psi}
\ee
This latter probability distribution is normalized to the unit value as
\be
\sum_{\bf m} \mu({\bf m},l) = 1 \, ,
\ee
because of the hierarchy of Eqs.~(\ref{sum-Psi-1}), (\ref{sum-Psi-2}),...

Now, taking the limit $q\to 0$ in Eq.~(\ref{G_q-eq}) for ${\boldsymbol{\mathsf G}}_q(l)$ and in the further equations for ${\boldsymbol{\mathsf G}}_q({\bf m},l)$, we get
\bea
&& 0= {\boldsymbol{\mathsf A}}_l\cdot\pmb{\Psi}_{l-1} 
+ \sum_{m_{l+1}} {\boldsymbol{\mathsf W}}_{-m_{l+1},l+1}\cdot\pmb{\Psi}_{l+1}(m_{l+1}) \nonumber\\
&&- {\boldsymbol{\mathsf A}}_{l+1}\cdot\pmb{\Psi}_{l}
- \sum_{m_{l}} {\boldsymbol{\mathsf W}}_{-m_{l},l}\cdot\pmb{\Psi}_{l}(m_{l}) 
+ {\boldsymbol{\mathsf W}}_{0,l}\cdot\pmb{\Psi}_{l} \, ,
\label{eq-Psi-0}
\eea
\bea
0 &=& {\boldsymbol{\mathsf W}}_{+m_l} \cdot\pmb{\Psi}_{l-1} + \sum_{m_{l+1}} {\boldsymbol{\mathsf W}}_{-m_{l+1},l+1} \cdot \pmb{\Psi}_{l+1}(m_lm_{l+1}) \nonumber\\
&&- \left({\boldsymbol{\mathsf A}}_{l+1} + {\boldsymbol{\mathsf W}}_{-m_l,l} - {\boldsymbol{\mathsf W}}_{0,l}\right)\cdot\pmb{\Psi}_{l}(m_l) \, , \label{eq-Psi-1}
\eea
\bea
0 &=& {\boldsymbol{\mathsf W}}_{+m_l} \cdot\pmb{\Psi}_{l-1}(m_{l-1}) \nonumber\\
&&+ \sum_{m_{l+1}} {\boldsymbol{\mathsf W}}_{-m_{l+1},l+1} \cdot \pmb{\Psi}_{l+1}(m_{l-1}m_lm_{l+1}) \nonumber\\
&&- \left({\boldsymbol{\mathsf A}}_{l+1} + {\boldsymbol{\mathsf W}}_{-m_l,l} - {\boldsymbol{\mathsf W}}_{0,l}\right)\cdot\pmb{\Psi}_{l}(m_{l-1}m_l) \, , \label{eq-Psi-2}
\eea
etc..., thus forming a hierarchy of coupled equations for $\pmb{\Psi}_l({\bf m})$.

Here, we observe that Eq.~(\ref{eq-Psi-0}) can be obtained by summing Eq.~(\ref{eq-Psi-1}) over the monomeric unit $m_l$.  Similarly, Eq.~(\ref{eq-Psi-1}) is given by summing Eq.~(\ref{eq-Psi-2}) over $m_{l-1}$, and so on for the next equations in this hierarchy. Equation~(\ref{eq-Psi-1}) as well as the following ones for $\pmb{\Psi}_l(m_l)$, $\pmb{\Psi}_l(m_{l-1}m_l)$,... have the same structure.  This observation suggests to find the solution of this hierarchy by factorization into matrix products as
\bea
&&\pmb{\Psi}_l(m_l) = {\boldsymbol{\mathsf Y}}_{m_l,l} \cdot \pmb{\Psi}_{l-1} \label{fact-Psi1}\\
&&\pmb{\Psi}_{l+1}(m_{l}m_{l+1}) = {\boldsymbol{\mathsf Y}}_{m_{l+1},l+1} \cdot {\boldsymbol{\mathsf Y}}_{m_l,l} \cdot \pmb{\Psi}_{l-1} \label{fact-Psi2}\\
&&\qquad\qquad\vdots \nonumber
\eea
in terms of $2\times 2$ matrices ${\boldsymbol{\mathsf Y}}_{m_l,l}$, each one associated with the monomeric unit $m_l$ at the location $l$ of the template. If this assumption is valid, the probabilities~(\ref{mu-Psi}) that the tip of the copolymer has the monomeric subsequence $m_{l-r+1}\cdots m_{l-1}m_l$ at the location $l$ of the template should be given by
\bea
&& \mu(m_{l-r+1}\cdots m_{l-1}m_l,l) \nonumber\\
&&= {\rm tr}({\boldsymbol{\mathsf Y}}_{m_l,l} \cdots{\boldsymbol{\mathsf Y}}_{m_{l-r+1},l-r+1}\cdot\pmb{\Psi}_{l-r})/({\rm tr}\, \pmb{\Psi}_l) \, .
\label{mu-Ys}
\eea

Next, in order to satisfy Eqs.~(\ref{sum-Psi-1}), (\ref{sum-Psi-2}),..., the following relation should hold between all the $2\times 2$ matrices $\pmb{\Psi}_l$ defined along the template,
\be
\pmb{\Psi}_l = {\boldsymbol{\mathsf R}}_{l}\cdot \pmb{\Psi}_{l-1}
\label{forward_iter}
\ee
with
\be
{\boldsymbol{\mathsf R}}_{l}\equiv \sum_{m_l} {\boldsymbol{\mathsf Y}}_{m_l,l} \, .
\label{R-dfn}
\ee
Moreover, the assumptions~(\ref{fact-Psi1}), (\ref{fact-Psi2}),... are substituted into Eqs.~(\ref{eq-Psi-1}), (\ref{eq-Psi-2}),... Since all of them have the same structure, they are satisfied at once if the matrices~${\boldsymbol{\mathsf Y}}_{m_l,l}$ obey the following relation,
\bea
0 &=& {\boldsymbol{\mathsf W}}_{+m_l,l} + \bigg( \sum_{m_{l+1}} {\boldsymbol{\mathsf W}}_{-m_{l+1},l+1} \cdot {\boldsymbol{\mathsf Y}}_{m_{l+1},l+1}\nonumber\\
&&\qquad  - {\boldsymbol{\mathsf A}}_{l+1} - {\boldsymbol{\mathsf W}}_{-m_l,l} + {\boldsymbol{\mathsf W}}_{0,l}\bigg)\cdot{\boldsymbol{\mathsf Y}}_{m_l,l} \, .\label{eq-X}
\eea
Introducing the $2\times 2$ matrices
\be
{\boldsymbol{\mathsf V}}_{l}\equiv {\boldsymbol{\mathsf A}}_{l+1} -\sum_{m_{l+1}} {\boldsymbol{\mathsf W}}_{-m_{l+1},l+1} \cdot {\boldsymbol{\mathsf Y}}_{m_{l+1},l+1} \, ,
\label{V-dfn}
\ee
Eq.~(\ref{eq-X}) gives the matrix~${\boldsymbol{\mathsf Y}}_{m_l,l}$ as
\be
{\boldsymbol{\mathsf Y}}_{m_l,l} = ({\boldsymbol{\mathsf V}}_{l} -{\boldsymbol{\mathsf W}}_{0,l} +{\boldsymbol{\mathsf W}}_{-m_l,l})^{-1}\cdot {\boldsymbol{\mathsf W}}_{+m_{l},l} \, ,
\label{eq-X-V}
\ee
which can be replaced back into Eq.~(\ref{V-dfn}) to finally obtain the {\it backward iteration}
\be
\boxed{
{\boldsymbol{\mathsf V}}_{l-1}= ({\boldsymbol{\mathsf V}}_l-{\boldsymbol{\mathsf W}}_{0,l})\cdot \sum_m ({\boldsymbol{\mathsf V}}_l-{\boldsymbol{\mathsf W}}_{0,l}+ {\boldsymbol{\mathsf W}}_{-m,l})^{-1} \cdot {\boldsymbol{\mathsf W}}_{+m,l} 
}
\label{eq-V}
\ee
along the template.  We note that this recurrence only involves the rates, which are known {\it a priori}.  Accordingly, Eq.~(\ref{eq-V}) provides the sequence of matrices $\{ {\boldsymbol{\mathsf V}}_l\}$, which determines the matrices~(\ref{eq-X-V}) and thus the matrices~(\ref{R-dfn}).  Once these latter are obtained, the {\it forward iteration}~(\ref{forward_iter}) can run along the template in order to find the sequence of matrices $\{ \pmb{\Psi}_l\}$.  Hence, all the probabilities~(\ref{mu-Ys}) are fully determined with this matrix product method.

In order to obtain the mean growth velocity, Eq.~(\ref{eq-V}) can be rewritten in the following form,
\be
{\boldsymbol{\mathsf V}}_{l-1} = ({\boldsymbol{\mathsf V}}_l - {\boldsymbol{\mathsf W}}_{0,l})\cdot {\boldsymbol{\mathsf R}}_l \, .
\ee
Multiplying this equation to the left-hand side of the matrix $\pmb{\Psi}_{l-1}$ and using Eq.~(\ref{forward_iter}) gives
\be
{\boldsymbol{\mathsf V}}_{l-1}\cdot\pmb{\Psi}_{l-1} = ({\boldsymbol{\mathsf V}}_l - {\boldsymbol{\mathsf W}}_{0,l})\cdot \pmb{\Psi}_l  \, .
\ee
Taking its trace and using the fact that ${\rm tr}({\boldsymbol{\mathsf W}}_{0,l}\cdot \pmb{\Psi}_l)=0$, which is a consequence of the structure of the matrices~(\ref{W0-dfn}) and~(\ref{Psi-dfn}), we have that
\be
{\rm tr}({\boldsymbol{\mathsf V}}_{l-1}\cdot\pmb{\Psi}_{l-1}) = {\rm tr}({\boldsymbol{\mathsf V}}_l \cdot \pmb{\Psi}_l) = C \, ,
\label{C-dfn}
\ee
holding for all $l$ with some constant $C$.  Going back to Eq.~(\ref{G_q-eq}), we can now consider its derivative with respect to the wave number $q$, using the expansions~(\ref{dispersion1}) and~(\ref{G_q-expansion}) in powers of $q$.  Taking the trace of the resulting equation eliminates the terms involving the matrix~${\boldsymbol{\mathsf W}}_{0,l}$.  Furthermore, we suppose that the template is a periodic sequence with a long period $L$, and we take the sum of the equations over $l=1,2,...,L$.  Since ${\boldsymbol{\mathsf G}}_0(l)=\pmb{\Psi}_l$, using Eq.~(\ref{V-dfn}), we finally obtain the mean growth velocity as
\be
\boxed{
v= \frac{\sum_{l=1}^L{\rm tr}({\boldsymbol{\mathsf V}}_{l}\cdot\pmb{\Psi}_l)}{\sum_{l=1}^L{\rm tr}\, \pmb{\Psi}_l}
}\label{v-V-Psi}
\ee
for a periodic template.  In the case of a uniform template (i.e., of period $L=1$), we recover the expression previously obtained in Ref.~\onlinecite{G19JCP} for multistate template-free copolymerization.

Since the relation~(\ref{C-dfn}) holds along the template, Eq.~(\ref{v-V-Psi}) implies
\be
v= \frac{LC}{\sum_{l=1}^L{\rm tr}\, \pmb{\Psi}_l} \, .
\ee
If we introduce the quantities
\be
\tau_l \equiv \frac{{\rm tr}\, \pmb{\Psi}_l}{{\rm tr}({\boldsymbol{\mathsf V}}_{l}\cdot\pmb{\Psi}_l)} = 
\frac{1}{C}\, {\rm tr}\, \pmb{\Psi}_l
\label{tau_l}
\ee
and take the limit of an arbitrarily long period $L$,
the mean growth velocity can thus be expressed as
\be
v= \frac{1}{\langle\tau_l\rangle}
\label{mean_v}
\ee
with the statistical average along the template defined by
\be
\langle\tau_l\rangle\equiv \lim_{L\to\infty} \frac{1}{L} \, \sum_{l=1}^L \tau_l \, .
\ee
The quantities~(\ref{tau_l}) have the interpretation of mean times spent by the growing copolymer at the location $l$ of the template.  In view of the result~(\ref{v-V-Psi}), the matrices~${\boldsymbol{\mathsf V}}_{l}$ can be interpreted as representing the local velocities and $\pmb{\Psi}_l$~determine the probabilities to find the length of the copolymer at the successive locations $l$ along the template.

These different local quantities characterize the effects due to the heterogeneity of the template.  If the template is random, it forms a disordered medium for the growth of the copolymer and anomalous effects similar to those existing for random walks in such media manifest themselves here as well.\cite{HL97,JB98,WEMO98,KLN04,BSW04,BFW07,WGM95,BAO78,BS83,DP82,D83,ABPS90,BCGL90} In particular, if the mean times~(\ref{tau_l}) have a probability distribution with an algebraic tail as $p(\tau)\sim 1/\tau^{\gamma+1}$ for $\tau\to\infty$, the growth of the copolymer is sublinear as $\langle l\rangle_t \sim t^{\gamma}$ if $0<\gamma<1$, but linear as $\langle l\rangle_t \simeq vt$ if $\gamma>1$.  However, template-directed copolymerization processes differ from standard random walks in disordered media by the reactions of attachment and detachment of monomers and the transitions between multiple activation states, which add complexity to these growth processes.\citep{G16PRL,G17JSM}

As aforementioned, the theory also applies to mechanisms with more than $I=2$ activation states by considering $I\times I$ matrices for any number $I$ of activation states.

\subsection{Thermodynamics}
\label{Thermo}

In the regime of steady growth at positive mean growth velocity $v>0$, the thermodynamic entropy production rate is given in general by\cite{AG08}
\be
\frac{1}{k_{\rm B}} \frac{d_{\rm i}S}{dt} = v \, \left[ \epsilon + D(\omega\vert\alpha)\right] \equiv v \, A \geq 0 \, ,
\label{entrprod}
\ee
where $\epsilon$ is the free-energy driving force, $D(\omega\vert\alpha)$ the conditional Shannon disorder of the copy $\omega$ with respect to the template $\alpha$, and $A=\epsilon+D(\omega\vert\alpha)$ the entropy production per monomeric unit, also called affinity.  On the one hand, the free-energy driving force is defined as $\epsilon=-g/T$ in terms of the free energy $g$ per monomeric unit and the temperature~$T$.\cite{AG08,B79}  On the other hand, the conditional Shannon disorder is determined by the probability distribution~(\ref{mu-Psi-i}) of the copy sequence according to\cite{AG08}
\be
D(\omega\vert\alpha) = \lim_{l\to\infty} -\frac{1}{l} \sum_{{\bf m},i}
\mu({\bf m},l,i)\, \ln\mu({\bf m},l,i) \, .
\label{D-gen}
\ee
The conditional Shannon disorder can be expressed as
\be
D(\omega\vert\alpha)=D(\omega)-I(\omega,\alpha) \geq 0
\label{MI}
\ee
in terms of the overall Shannon disorder $D(\omega)$ of the copy and the mutual information $I(\omega,\alpha)$ between the copy and the template.\cite{CT06}  The conditional Shannon disorder characterizes the amount of replication errors, and the mutual information the replication fidelity.  In the absence of replication errors, we would have that $D(\omega\vert\alpha)=0$ and $I(\omega,\alpha)=D(\alpha)=D(\omega)=-\sum_{n=1}^M \nu_n\ln \nu_n$, but this would require the vanishing of the attachment rates for incorrect pairs $m_l:n_l$ between the copy $\omega$ and the template $\alpha$.

\section{Single-state mechanism for template-directed copolymerization}
\label{1state}

\subsection{Iterated function system}

In the case where there is a single activation state ($I=1$), the transition rates $w_l^{i\to j}$ are vanishing, so that ${\boldsymbol{\mathsf W}}_{0,l}=0$.  Accordingly, all the matrices can be reduced to scalars.  In particular, the matrices ${\boldsymbol{\mathsf W}}_{\pm m_l,l}$ are reduced to $w_{\pm m_l,l}$, and 
${\boldsymbol{\mathsf V}}_{l}$ to $v_l$.  Consequently, the backward iteration~(\ref{eq-V}) becomes
\be
v_{l-1} = v_l \sum_{m=1}^{M} \frac{w_{+m,l}}{v_l+w_{-m,l}} \, .
\label{v_l-iteration}
\ee
Moreover, $\pmb{\Psi}_l$ reduces to the scalar $\psi_l$ and Eq.~(\ref{C-dfn}) implies that $v_l\psi_l=C$ for all integers $l$.  If we define
\be
x_l\equiv v_l=\frac{C}{\psi_l} = \frac{1}{\tau_l}
\ee
the mean growth velocity~(\ref{mean_v}) is given by
\be
\frac{1}{v} = \left\langle\frac{1}{x_l}\right\rangle \, ,
\ee
as in previous work.\cite{G16PRL,G17JSM}

If the attachment and detachment rates only depend on the unit $n_l$ at the location $l$ of the template,
$w_{\pm m_l,l}=w_{\pm m_l,n_l}$, the backward iteration~(\ref{v_l-iteration}) can be expressed as the following {\it iterated function system} (IFS),
\be
x_{l-1} = f_{n_l}(x_l) \qquad\mbox{with} \qquad
f_n(x) \equiv x \sum_{m=1}^M \frac{w_{+m,n}}{x+w_{-m,n}}\, ,
\label{f_n}
\ee
for $n=1,2,...,M$.  During the backward iteration along the template, the function $f_{n_l}$ corresponding to the unit $n_l$ of the template is used to obtain the local velocity $x_{l-1}$ from $x_l$.  If the template is a Bernoulli chain of probabilities $(\nu_1,\nu_2,...,\nu_M)$, these functions are randomly picked according to these probabilities as the backward iteration is run.  Once the local velocities $\{x_l\}$ are known, the probabilities to find the monomeric unit $m_l$ at the location $l$ of the template are given by
\be
\mu(m_l, l) = \frac{x_l}{x_{l-1}} \frac{w_{+m_l , n_l}}{x_l+w_{-m_l , n_l}} \, ,
\label{mu-B}
\ee
which is deduced from Eqs.~(\ref{mu-Psi}), (\ref{fact-Psi1}), and~(\ref{eq-X-V}) for a single activation state $I=1$.  Consequently, the probability~(\ref{mu-Ys}) to find some subsequence is factorized as
\be
\mu(m_{l-r+1} \cdots m_{l-1}m_l ,l) = \prod_{k=l-r+1}^{l}  \mu(m_k, k) \, ,
\label{proba-B}
\ee
so that the sequence of the copolymer is a Bernoulli chain, which depends on the location along the template.  Nevertheless, there is no correlation between the successive monomeric units composing the copolymer.  The reasons are that there is a single activation state and the rates only depend on the ultimate monomeric unit that is currently attached or detached in the kinetic scheme that is here considered.  If the rates also depended on previously incorporated monomeric units, the copolymer would form a Markov chain.\cite{G17JSM,G17PRE,GA14,G16,SSOL17}  Equation~(\ref{mu-B}) gives the pairing probabilities $\mu(m_l:n_l)=\mu(m_l,l)$ between the monomeric units of the copy and the template and, in particular, the probabilities of correct and incorrect pairs, which determine the replication errors.\cite{G16PRL,G17JSM,G17PRE}

As previously shown,\cite{G16PRL,G17JSM} the exponent $\gamma$ controlling the probability distribution of the mean times~$\tau_l=1/x_l$ is obtained with the following relation,
\be
\sum_{n=1}^M \frac{\nu_n}{f'_n(0)^{\gamma}} = 1 \, ,
\label{gamma_eq}
\ee
where
\be
f'_n(0) = \sum_{m=1}^M \frac{w_{+m,n}}{w_{-m,n}} \, .
\label{f'}
\ee
The relation~(\ref{gamma_eq}) allows us to determine the domain of parameter values where the growth is sublinear in time (if $0<\gamma<1)$ or linear (if $\gamma>1$).  The growth is essentially stopped if the exponent vanishes, $\gamma=0$.  Expanding $f'_n(0)^{-\gamma}=\exp[-\gamma f'_n(0)]$ in powers of $\gamma$ in Eq.~(\ref{gamma_eq}), the equation is satisfied for $\gamma=0$ under the condition that
\be
\sum_{n=1}^M \nu_n \ln f'_n(0) = 0 \qquad\mbox{or}\qquad \prod_{n=1}^M f'_n(0)^{\nu_n} = 1 \, .
\label{gamma=0}
\ee
In this regime, the distribution of local velocities $x_l$ is concentrated at  $x=0$.

For copolymers forming the Bernoulli chains~(\ref{proba-B}), the thermodynamic entropy production rate is given by Eq.~(\ref{entrprod}) with the free-energy driving force and the conditional Shannon disorder respectively evaluated by
\be
\epsilon \equiv \lim_{L\to\infty} \frac{1}{L} \sum_{l=1}^L \sum_{m_l=1}^{M}
\mu(m_l,l)\, \ln\frac{w_{+m_l,n_l}}{w_{-m_l,n_l}} \, ,
\label{eps-B}
\ee
and
\be
D(\omega\vert\alpha) \equiv \lim_{L\to\infty} -\frac{1}{L} \sum_{l=1}^L \sum_{m_l=1}^{M}
\mu(m_l,l)\, \ln\mu(m_l,l) \, ,
\label{D-B}
\ee
both per monomeric unit.\cite{G17JSM}

Thermodynamic equilibrium can be identified as happening for the conditions where detailed balance is satisfied and the entropy production rate~(\ref{entrprod}) is vanishing.  Under these conditions, both the mean growth velocity and the affinity are equal to zero.  These conditions are implied by the vanishing of the exponent $\gamma$ according to Eq.~(\ref{gamma=0}).  Indeed, substituting Eq.~(\ref{mu-B}) in Eqs.~(\ref{eps-B}) and~(\ref{D-B}), eliminating the ratio $x_l/x_{l-1}$ by using Eq.~(\ref{f_n}), and taking the limit of vanishing local velocities $x_l\simeq 0$, the affinity becomes $A=\epsilon+D(\omega\vert\alpha)=\lim_{L\to\infty} (1/L) \sum_{l=1}^L \ln f'_{n_l}(0)=\sum_{n=1}^M \nu_n \ln f'_n(0)$, so that $A=0$ if Eq.~(\ref{gamma=0}) holds, i.e., if $\gamma=0$.

\subsection{Illustrative example}

We consider template-directed copolymerization with two monomeric species ($M=2$).  The rates are supposed to obey the mass-action law, so that the attachment rates are proportional to the concentrations $(c_1,c_2)$ of monomers in the surrounding solution, while the detachment rates are not:
\be
w_{+m,n} = k_{+m,n}\, c_m \, ,\qquad w_{-m,n} = k_{-m,n} \, ,
\ee
where $k_{\pm m,n}$ denote the rate constants.  These latter are taken as
\bea
\label{SET2}
&& k_{+1,1}=0.5\, , \ k_{+1,2}=0.3 \, ,  \ k_{+2,1}=0.2 \, , \  k_{+2,2}=2 \, ,   \nonumber \\
&& k_{-1,1}=0.1 \, , \ k_{-1,2}=0.2  \, , \ k_{-2,1}=0.3  \, , \ k_{-2,2}=0.4 \, . \nonumber\\
&& 
\eea
Moreover, the template is assumed to be a Bernoulli chain of probabilities $\nu_1=\nu_2=0.5$, except if otherwise said.  

\begin{figure}[h]
\centerline{\scalebox{0.65}{\includegraphics{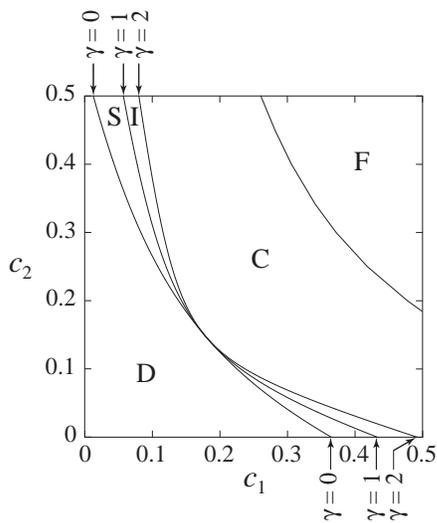}}}
\caption{Single-state templated-directed copolymerization with the parameter values~(\ref{SET2}) yielding Bernoulli chains: Space of the monomeric concentrations $c_1$ and $c_2$.  The template is a Bernoulli chain of probabilities $(0.5,0.5)$.  See the text for explanation.}
\label{fig1}
\end{figure}

Figure~\ref{fig1} shows the different regimes of template-directed copolymerization in the plane of the concentrations $(c_1,c_2)$ for both types of monomers.  The lines with the exponent values $\gamma=0,1,2$ are obtained with Eqs.~(\ref{gamma_eq}) and~(\ref{f'}) for the present example.  Thermodynamic equilibrium happens along the line $\gamma=0$ where the growth is stopped together with the entropy production per monomer, $A=0$.  In the domain D below the line $\gamma=0$, depolymerization occurs for a copolymer initially present in the solution, because the concentrations are too low for attachment to dominate over detachment and growth to occur.  In the domain S between the lines $\gamma=0$ and $\gamma=1$, the growth is sublinear in time as $\langle l \rangle_t\sim t^{\gamma}$.  In the intermediate domain I between the lines $\gamma=1$ and $\gamma=2$, the growth is linear in time with a positive mean growth velocity $v>0$, but the diffusivity is infinite ${\cal D}=\infty$.  In the domain C beyond the line $\gamma=2$, the mean growth velocity is positive $v>0$ and the diffusivity is positive and  finite $0<{\cal D}<\infty$.  The distribution of local velocities $\{x_l\}$ is continuous in the domain C, but it becomes fractal in the domain F.

\begin{figure}[h]
\centerline{\scalebox{0.495}{\includegraphics{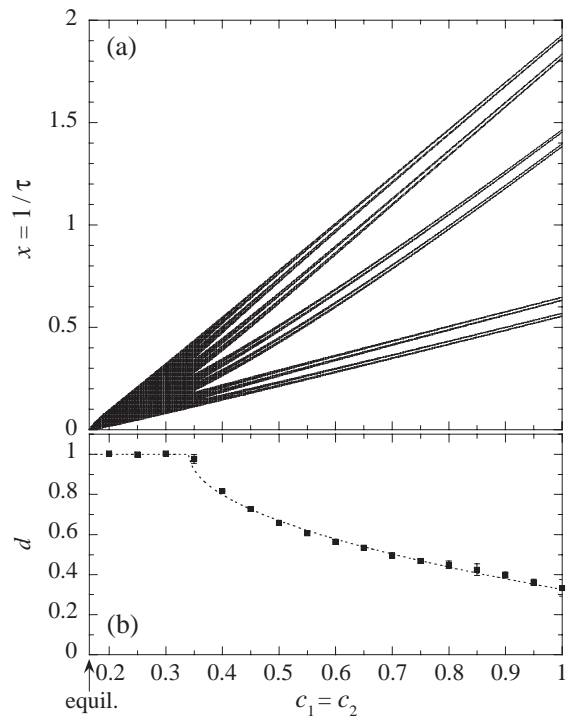}}}
\caption{Single-state templated-directed copolymerization with the parameter values~(\ref{SET2}) yielding Bernoulli chains for the concentrations $c_1=c_2$: (a) Invariant set of the IFS~(\ref{f_n}) for the local velocity $x=x_l=1/\tau_l$ versus the common concentration $c_1=c_2$ of the monomers.  (b) Box counting dimension $d$. For every value of concentration, the distribution is generated with the IFS running along a template sequence of length $10^6$ and taken as a Bernoulli chain of probabilities $(0.5,0.5)$.}
\label{fig2}
\end{figure}

This fractal is observed in Fig.~\ref{fig2}(a) depicting the distribution of the local velocities $\{x_l\}$ as a function of equal concentrations $c_1=c_2$ for both monomers.  Gaps appear in the distribution of local velocities as the concentrations increase.  This happens at the critical concentration value where the two functions~(\ref{f_n}) coincide as $f_1(x_2)=f_2(x_1)$ for the fixed points of each other.  These fixed points are given by $x_n=f_n(x_n)$ for $n=1,2$.  Beyond this critical value, the dimension of the distribution is smaller than one and the distribution becomes fractal, as seen in Fig.~\ref{fig2}(b).

For equal concentrations $c_1=c_2$, the critical concentration values corresponding to the transitions between the different regimes are the following:
\bea
\gamma=0:&& \quad c_1=c_2=0.16477 \, , \\
\gamma=1:&& \quad c_1=c_2=0.16516 \, , \\
\gamma=2:&& \quad c_1=c_2=0.16555 \, , \\
\mbox{C-F transition}:&& \quad c_1=c_2=0.34130 \, .
\eea

\begin{figure}[h]
\centerline{\scalebox{0.495}{\includegraphics{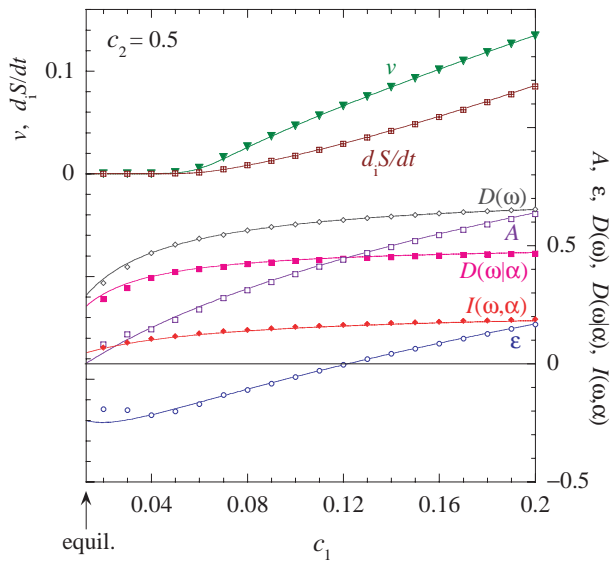}}}
\caption{Single-state templated-directed copolymerization with the parameter values~(\ref{SET2}) yielding Bernoulli chains for the concentration $c_2=0.5$:  Mean growth velocity $v$ (filled triangles), affinity $A$ (open squares), free-energy driving force $\epsilon$ (open circles), conditional Shannon disorder $D(\omega\vert\alpha)$ (filled squares), overall Shannon disorder $D(\omega)$ (open diamonds), mutual information $I(\omega,\alpha)$ (filled diamonds), and entropy production rate $d_{\rm i}S/dt$ (crossed squares) versus the concentration $c_1$.  The data points are obtained by statistics over a sample of $10^2$ copy sequences grown by Monte Carlo simulations along a template of length $10^6$.  The lines are the theoretical results obtained with the IFS~(\ref{f_n}) running along a template sequence of length $10^5$ and taken as Bernoulli chain of probabilities $(0.5,0.5)$.}
\label{fig3}
\end{figure}

In Fig.~\ref{fig3}, the mean growth velocity and the different thermodynamic quantities are shown for increasing values of the concentration $c_1$ at $c_2=0.5$.  The data points are the results of kinetic Monte Carlo simulations with Gillespie's algorithm,\cite{G76,G77} while the lines are obtained with the iterated function system~(\ref{f_n}).  The coincidence of the lines with the data points is numerical evidence for the exactness of theory in solving the kinetic equations.  For the data points at the two smallest values of the concentration ($c_1=0.02,0.03$), the growth is sublinear and extremely slow so that the copolymer sequence has not grown beyond a length of $10^3$ even after $10^8$ steps of the Monte Carlo method, resulting into very poor statistics, which explains the deviations observed for these data points.  In any case, the method based on the iterated function system is computationally much faster than the kinetic Monte Carlo method.  The gain factor in computational time is about $10^6$ in order to get a comparable accuracy.

The critical values of the concentration $c_1$ corresponding to the different transitions in Fig.~\ref{fig3} are given by
\bea
\gamma=0:&& \quad c_2=0.5 \, , \quad c_1=0.01273 \, , \\
\gamma=1:&& \quad c_2=0.5 \, , \quad c_1=0.05730 \, , \\
\gamma=2:&& \quad c_2=0.5 \, , \quad c_1=0.08020 \, , \\
\mbox{C-F transition}:&& \quad c_2=0.5 \, , \quad c_1=0.26068 \, .
\eea
We see in this figure that equilibrium indeed happens at $c_1=0.01273$ since the affinity is vanishing at this concentration value, $A=\epsilon+D(\omega\vert\alpha)=0$. At equilibrium, the free-energy driving force is thus negative and given by $\epsilon=-D(\omega\vert\alpha)$ in terms of the equilibrium value of the conditional Shannon disorder: $D(\omega\vert\alpha)\simeq 0.24$.  Moreover, the mean growth velocity and the entropy production rate are equal to zero between equilibrium at $\gamma=0$ and the threshold of positive velocity at $\gamma=1$, i.e., for $0.01273\leq c_1\leq 0.05730$.  For $c_1>0.05730$, the growth is linear in time with a positive mean growth velocity.  In this illustrative example, a lot of replication errors occur because the conditional Shannon disorder takes a relatively large value with respect to its upper bound $D(\omega\vert\alpha)\le \ln2=0.69315$ as seen in Fig.~\ref{fig3}, while the mutual information between the copy and the template is much smaller than its upper bound $I(\omega,\alpha)\le \ln2=0.69315$, meaning that replication fidelity is low.  The free-energy driving force is negative $\epsilon<0$ for $0.01273<c_1\leq 0.122$.  In this concentration range, the growth is powered by the entropic effect of replication errors, since the affinity is positive $A=\epsilon+D(\omega\vert\alpha)>0$, the magnitude of the conditional Shannon disorder compensating the unfavorable free-energy driving force.\cite{AG08}

\begin{figure}[h]
\centerline{\scalebox{0.4}{\includegraphics{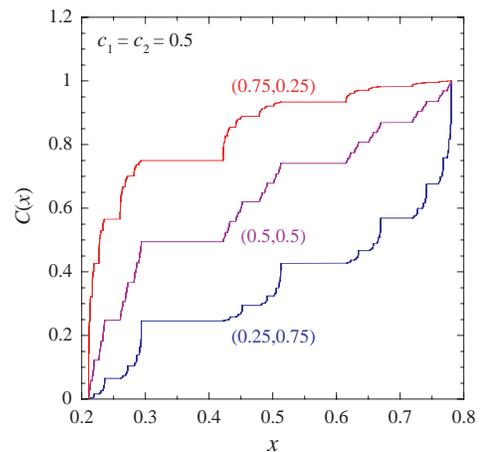}}}
\caption{Single-state templated-directed copolymerization with the parameter values~(\ref{SET2}) yielding Bernoulli chains and the concentrations $c_1=c_2=0.5$:  Cumulative functions $C(x)$ of the probability distribution for the local velocity $x=x_l$ along templates taken as Bernoulli chains of length $1.2\times 10^4$ and probabilities $(\nu_1,\nu_2=1-\nu_1)$ with the values $\nu_1=0.25,0.5,$ and $0.75$.}
\label{fig4}
\end{figure}

Figure~\ref{fig4} depicts the cumulative functions of the local velocity distribution $\{x_l\}$ at the concentrations $c_1=c_2=0.5$ for three different templates, respectively taken as Bernoulli chains of probabilities $(\nu_1,\nu_2)$ with $\nu_1=1-\nu_2=0.25$, $0.5$, and $0.75$.  Therein, we observe not only the horizontal plateaus corresponding to gaps in the fractal distribution, but also the differences in the probability weights given to each value of the local velocity $x$.  The plateau extending over the interval $0.294<x<0.422$ and corresponding to the main gap of the fractal takes the value $\nu_1$, which is the probability of the monomeric unit of type $1$ in the template.  Since the attachment rates $k_{+m,1}$ are globally smaller than the rates $k_{+m,2}$ for the other type $2$ in the parameter set~(\ref{SET2}), a larger probability $\nu_1$ will increase the probability weight at lower values of the local velocity $x$, so that the cumulative function for $\nu_1=0.75$ is larger than those for $\nu_1=0.25$ and $\nu_1=0.5$.  As a further consequence, the mean growth velocity should decrease as $\nu_1$ increases, which is confirmed by the following values of the mean growth velocity,
\be
\begin{array}{ll}
(\nu_1,\nu_2)=(0.25,0.75):& \quad v= 0.48074 \, , \\
(\nu_1,\nu_2)=(0.5,0.5):& \quad v= 0.34251 \, , \\
(\nu_1,\nu_2)=(0.75,0.25):& \quad v= 0.26243 \, ,
\end{array}
\ee
in the conditions of Fig.~\ref{fig4}.  These effects are caused by the heterogeneity of the template.

\section{Multistate mechanism for template-directed copolymerization}
\label{Multistate}

\subsection{Kinetics depending on the ultimate monomeric unit}

Here, we consider a multistate mechanism with two activation states ($I=2$). For this mechanism, the kinetic equations~(\ref{kin_eq}) can be solved with the method presented in Sec.~\ref{Kinetics}.  We suppose that the rates only depend on the ultimate monomeric unit $m_l$, as well as on the corresponding template monomeric unit $n_l$, with which the pairing $m_l:n_l$ may happen, so that $w_{\pm m_l,l}^{i}=w_{\pm m_l,n_l}^{i}$ and $w_{l}^{i\to j}=w_{n_l}^{i\to j}$.  These rates enter into the definitions of the matrices~(\ref{W0-dfn}) and~(\ref{W-dfn}).  The backward iteration~(\ref{eq-V}) here forms the {\it iterated matrix function system} (IMFS):
\be
{\boldsymbol{\mathsf V}}_{l-1}= {\boldsymbol{\mathsf F}}_{n_l}({\boldsymbol{\mathsf V}}_l)
\label{IMFS1}
\ee
with the matrix functions
\be
{\boldsymbol{\mathsf F}}_n({\boldsymbol{\mathsf V}})= ({\boldsymbol{\mathsf V}}-{\boldsymbol{\mathsf W}}_{0,n})\cdot \sum_m ({\boldsymbol{\mathsf V}}-{\boldsymbol{\mathsf W}}_{0,n}+ {\boldsymbol{\mathsf W}}_{-m,n})^{-1} \cdot {\boldsymbol{\mathsf W}}_{+m,n} 
\label{IMFS2}
\ee
for $n=1,2,...,M$.  This iteration is run backward along the template to obtain the local velocity matrices $\{{\boldsymbol{\mathsf V}}_l\}$.  Their knowledge gives the matrices $\{{\boldsymbol{\mathsf R}}_l\}$ according to Eqs.~(\ref{R-dfn}) and~(\ref{eq-X-V}).  Therefore, the forward iteration~(\ref{forward_iter}) can be run to provide the matrices $\{\pmb{\Psi}_l\}$, which are combined with the local velocity matrices to obtain the mean growth velocity according to Eq.~(\ref{v-V-Psi}).  In order to converge numerically, the backward and forward iterations can run several times cyclically along a long template of length $L$.  Here below, these theoretical results are compared with kinetic Monte Carlo simulations using Gillespie's algorithm.\cite{G76,G77}

The knowledge of the local velocity matrices~$\{{\boldsymbol{\mathsf V}}_l\}$ determines also the matrices~(\ref{eq-X-V}) and thus the matrix products giving the probabilities~(\ref{mu-Ys}) of the copolymer sequences.  Because of the transitions between the activation states, these probabilities can no longer be factorized as for Bernoulli chains in Eq.~(\ref{proba-B}).  Thus, the matrix products and the trace in Eq.~(\ref{mu-Ys}) have for consequence that the copolymer sequences form non-Markovian chains because of the multistate kinetics.\cite{CF63JCP,G19JCP,AG09JCP}

Here, the thermodynamic entropy production rate is given by Eq.~(\ref{entrprod}) and can be evaluated as well.\cite{G19JCP}  Thermodynamic equilibrium can be identified with the detailed balance conditions as explained in Appendix~\ref{AppA}.

\subsection{Illustrative example}

As in Sec.~\ref{1state}, we assume that the rates obey the mass-action law and that there are two types of monomeric units ($M=2$).  The attachment rates are thus proportional to the monomeric concentrations $c_m$ in the solution, but neither the detachment rates nor the transition rates do depend on the concentrations:
\bea
&&w_{+m,n}^{i}=k_{+m,n}^{i} c_m \, , \qquad w_{-m,n}^{i}=k_{-m,n}^{i} \, , \qquad\mbox{and}\nonumber\\
&& w_{n}^{i\to j}=k_{n}^{i\to j} \, ,
\label{model-rates}
\eea
for $m,n=1,2$ and $i,j=1,2$.  We take the following set of parameters,
\bea
&& k_{+1,1}^1=2 \, , \quad k_{+2,1}^1=4 \, , \quad k_{+1,1}^2=4 \, , \quad k_{+2,1}^2=2 \, , \nonumber\\
&& k_{-1,1}^1=1 \, , \quad k_{-2,1}^1=6 \, , \quad k_{-1,1}^2=2 \, , \quad k_{-2,1}^2=3 \, , \nonumber\\
&& k_{+1,2}^1=1 \, , \quad k_{+2,2}^1=2 \, , \quad k_{+1,2}^2=2 \, , \quad k_{+2,2}^2=1 \, , \nonumber\\
&& k_{-1,2}^1=3 \, , \quad k_{-2,2}^1=4 \, , \quad k_{-1,2}^2=6 \, , \quad k_{-2,2}^2=2 \, , \nonumber\\
&& k_1^{1\to 2}=1 \, , \quad k_2^{1\to 2}=2 \, , \quad k_1^{2\to 1} = 2 \, , \quad k_2^{2\to 1} = 4 \, , \nonumber\\
&&c_{\rm 2}=1 \, , \label{ExM}
\eea
which satisfy the conditions for the existence of equilibrium as explained in Appendix~\ref{AppA}.

\begin{figure}[h]
\centerline{\scalebox{0.495}{\includegraphics{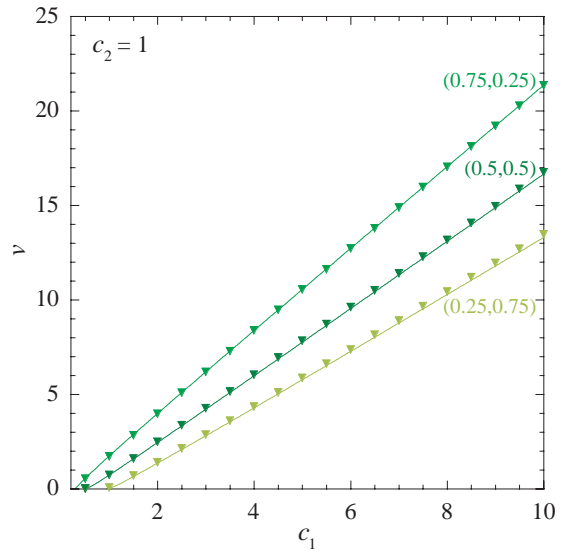}}}
\caption{Multistate templated-directed copolymerization with the parameter values~(\ref{ExM}): The mean growth velocity~(\ref{mean_v}) versus the concentration $c_1$ along templates taken as Bernoulli chains of probabilities $(\nu_1,\nu_2=1-\nu_1)$ with the values $\nu_1=0.25,0.5$, and $0.75$. The data points are obtained by statistics over a sample of $10^2$ copy sequences grown by Monte Carlo simulations along a template of length $10^5$.  The lines are the theoretical results obtained with the IMFS~(\ref{eq-V}) running along template sequences of length $10^3$.}
\label{fig5}
\end{figure}

Figure~\ref{fig5} shows the mean growth velocity versus the concentration $c_1$ for copolymers grown along three kinds of templates taken as Bernoulli chains of probabilities $(\nu_1,\nu_2)$ with $\nu_1=1-\nu_2=0.25,0.5$, and $0.75$.  Here, the velocity is larger if the template contains more units of type $1$ since the attachment rates are larger for this type of template unit than for the other one in the parameter set~(\ref{ExM}).  Again, there is excellent agreement between the results of kinetic Monte Carlo simulations (data points) and the lines giving the predictions of the iterated matrix function system~(\ref{IMFS1})-(\ref{IMFS2}).

\begin{figure}[h]
\centerline{\scalebox{0.495}{\includegraphics{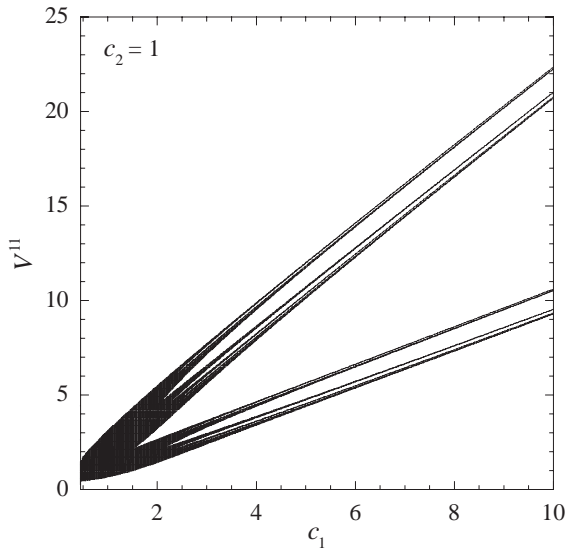}}}
\caption{Multistate templated-directed copolymerization with the parameter values~(\ref{ExM}): Invariant distribution of the local velocity matrix element $({\boldsymbol{\mathsf V}}_l)^{ij}$ for $i=j=1$, as a function of the concentration~$c_1$.  For every value of this concentration, the distribution is generated with the IMFS~(\ref{eq-V}) running along a template sequence of length $10^4$ and taken as a Bernoulli chain of probabilities $\nu_1=\nu_2=0.5$.}
\label{fig6}
\end{figure}

Here also, the backward iteration generates a fractal distribution of the local velocities.  Since the iteration is here carried out with matrices, we should consider the matrix elements of the local velocity matrices.  Figure~\ref{fig6} depicts the distribution of the elements $\{V_l^{11}\}$, showing that the fractal character appears beyond some critical value for the concentration $c_1$, as for the single-state mechanism of Sec.~\ref{1state}.

\begin{figure}[h]
\centerline{\scalebox{0.495}{\includegraphics{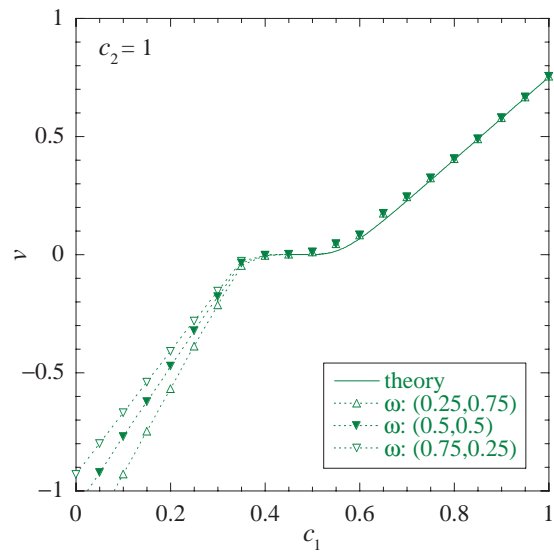}}}
\caption{Multistate templated-directed copolymerization with the parameter values~(\ref{ExM}): Mean velocity versus the concentration $c_1$ along a template taken as a Bernoulli chain of probabilities $\nu_1=\nu_2=0.5$, for an initial copy given by Bernoulli chains of probabilities $(\mu_1,\mu_2=1-\mu_1)$ with the values $\mu_1=0.25,0.5$, and $0.75$.  The data points are obtained by statistics over a sample of $10^2$ copy sequences grown by Monte Carlo simulations along a template of length $10^6$, the initial copy sequence having half the template length.  The solid line at positive velocities depicts the theoretical result for the mean growth velocity~(\ref{mean_v}) obtained with the IMFS~(\ref{eq-V}) running along a template of length $10^5$.  The dashed lines at negative velocities are only joining the data points of Monte Carlo simulations.}
\label{fig7}
\end{figure}

Figure~\ref{fig7} gives evidence for sublinear growth close to equilibrium in the multistate mechanism, as it is also the case in the single-case mechanism.\cite{G16PRL}  This figure shows the results of kinetic Monte Carlo simulations along a template~$\alpha$ of length~$10^6$, starting with an initial copolymer~$\omega$ paired with the template~$\alpha$ from $l=1$ up to $l=5\times 10^5$.  The sequence of the initial copolymer~$\omega$ is taken as a Bernoulli chain of probabilities $(\mu_1,\mu_2)$ with $\mu_1=1-\mu_2=0.25,0.5$, and~$0.75$.  If the concentration~$c_1$ is large enough, the copolymer $\omega$ extends beyond its initial length $l=5\times 10^5$ at the mean growth velocity~(\ref{mean_v}) predicted by theory and depicted by the solid line.  However, if the concentration~$c_1$ is too low, the initial copolymer undergoes depolymerization and its length decreases at some negative velocity.  This depolymerization velocity is not predicted by the growth theory and it depends on the initial composition $(\mu_1,\mu_2)$ of the copolymer chain that depolymerizes, as seen in Fig.~\ref{fig7}.  Now, this figure also shows that there is a range of concentration values where the mean velocity is vanishing, because the mean length of copolymerization or depolymerization depends sublinearly on time.  This anomalous behavior is a feature common with random walks in disordered media and with single-state template-directed copolymerization.\cite{G16PRL}  This anomalous behavior, as well as the fractal distribution of local velocities are two effects due to the heterogeneity of the template.

\section{Conclusion}
\label{Conclusion}

The template-directed growth of copolymers plays a central role for information processing at the molecular scale in biological cells.  In this paper, the theory of multistate template-directed reversible copolymerizations is developed by combining the iterated function systems previously obtained for single-state template-directed reversible copolymerization\cite{G16PRL,G17JSM} with the matrix product method pioneered by Coleman and Fox for multistate template-free irreversible polymerizations\cite{CF63JCP,CF63JACS,CF63JPS} and extended to multistate template-free reversible copolymerizations in Ref.~\onlinecite{G19JCP}.  

The theory here presented shows that the mean growth velocity, as well as the probabilities of the growing copolymer sequences can be obtained using an {\it iterated matrix function system} running backward along the template and complemented with a forward iteration.  This method provides the exact solution of the kinetic equations in the long-time limit and it is computationally much faster than kinetic Monte Carlo simulations.  The sequence probabilities are given in terms of matrix products, so that the grown copolymers form non-Markovian chains of monomeric units.

This theory gives a detailed understanding of the effects due to the heterogeneity of the template sequence on the growth of the copolymer.  In particular, this growth proceeds with a local velocity, which varies from location to location along the template.  Moreover, the probabilities of the copolymer sequences also depend on the template sequence in a way that is determined by matrix products obtained from the iterated matrix function system.  The growth process is thus controlled by the specificity and the heterogeneity of the template sequence.  As a consequence, the local velocity turns out to have a fractal distribution in regimes far enough from equilibrium.  As another consequence of template heterogeneity, the growth becomes anomalous in a regime close to equilibrium where the mean growth velocity vanishes in a finite interval of concentrations.  In this regime close to equilibrium, the growth is no longer linear, but sublinear in time.  

These effects can manifest themselves in biological processes such as DNA replication described as single-state templated-directed reversible copolymerization.\cite{G16PRL,G17PRE}  The generalization of such studies to multistate processes is important since DNA polymerases are known to undergo conformational changes between different structural states\cite{TJ06} and will be investigated in future work.  The study of template-directed copolymerization beyond the low conversion conditions is also of great importance, in particular, to understand the emergence of replicating molecular species in prebiotic chemistry.\cite{BL17,TB19}

Moreover, we may wonder if effects similar to those observed in the template-directed growth of copolymers would not also manifest themselves in other growth processes of spatial structures with more than one dimension.

\begin{acknowledgments}
This research is financially supported by the Universit\'e libre de Bruxelles (ULB) and the Fonds de la Recherche Scientifique~-~FNRS under the Grant PDR~T.0094.16 for the project ``SYMSTATPHYS".
\end{acknowledgments}

\appendix

\section{Equilibrium conditions for the multistate mechanism}
\label{AppA}

In order to identify a possible state of thermodynamic equilibrium for multistate template-directed reversible copolymerization ruled by the coupled master equations~(\ref{kin_eq}) with the rates~(\ref{model-rates}), the following detailed balance conditions should hold,
\bea
&& w_{+m_l,n_l}^{i} \, P_{\rm eq}(m_1\cdots m_{l-1}, l-1,i) \nonumber\\
&&\qquad\qquad\qquad = w_{-m_l,n_l}^{i} \, P_{\rm eq}(m_1\cdots m_{l-1}m_l,l,i) , \label{det-bal-1}\\
&& w_{n_l}^{i\to j} \, P_{\rm eq}(m_1\cdots m_l,l,i) = w_{n_l}^{j\to i}  \, P_{\rm eq}(m_1\cdots m_l,l,j) . \qquad \label{det-bal-2}
\eea
This is in particular the case if the transition rates satisfy
\be
\frac{w_{n}^{i\to j}}{w_{n}^{j\to i}} = K_{\rm eq}^{i\rightleftharpoons j} \qquad\mbox{for}\qquad n=1,2 \, ,
\label{cond-1}
\ee
and, moreover, if
\be
\frac{w_{+m,n}^{1}}{w_{-m,n}^{1}} = \frac{w_{+m,n}^{2}}{w_{-m,n}^{2}} \qquad\mbox{for}\qquad m,n=1,2 \, .
\label{eq-w-equil}
\ee
These conditions are satisfied for the parameter set~(\ref{ExM}).


\end{document}